\documentclass{aa}  

\usepackage{graphicx}
\usepackage{txfonts}
\usepackage{hyperref}
\usepackage{newtxtext}
\usepackage[varvw]{newtxmath}

\usepackage{amssymb}
\usepackage{lipsum}
\usepackage{afterpage}
\usepackage{threeparttable}
\usepackage{placeins}
\usepackage{multirow,makecell,booktabs}
\usepackage{graphicx}
\usepackage{subfig}
\usepackage[nameinlink]{cleveref}

\Crefname{figure}{Fig.}{Figs.}
\Crefname{equation}{Eq.}{Eqs.}
\Crefname{section}{Sect.}{Sects.}
\usepackage{enumitem}

\defcitealias{Sorgho2024}{S24}

\newcommand{\hi}{\textsc{Hi}}

\newcommand{\Mo}{M_{\odot}}
\newcommand{\Ms}{{M_*}}
\newcommand{\js}{$j_{\rm *}$}
\newcommand{\Mhi}{{M_{\textsc{Hi}}}}

\newcommand{\kms}{km\,s$^{-1}$}

\newcommand{\e}[1]{\times 10^{#1}}
\newcommand{\wise}{{\it WISE}}

\hypersetup{
    colorlinks = true,
    citecolor = blue,
    linkcolor = blue
}

\begin{document} 

\title{Angular momentum in isolated disc galaxies: Insights from TNG100}
\titlerunning{Angular momentum of Isolated discs}

\author{A. Sorgho\inst{1},
      L. Verdes-Montenegro\inst{1},
      M. Baes\inst{2},
      R. Ianjamasimanana\inst{1},
      M. Korsaga\inst{1,3},
      B. Namumba\inst{1},
      S. Sanchez-Exp\'osito\inst{1},
      J. Garrido\inst{1}
      }

\institute{\inst{1} Instituto de Astrof\'isica de Andaluc\'ia (CSIC), Glorieta de la Astronom\'ia s/n, 18008 Granada, Spain\\
          \email{asorgho@iaa.es}\\
          \inst{2} Sterrenkundig Observatorium, Universiteit Gent, Krijgslaan 281 S9, B-9000 Gent, Belgium\\
          \inst{3} Laboratoire de Physique et de Chimie de l'Environnement, Observatoire d'Astrophysique de l’Universit\'e Joseph Ki-Zerbo (ODAUO), 03 BP 7021, Ouaga 03, Burkina Faso
         }
\authorrunning{Sorgho et al.}

\date{Received xxx / Accepted xxx}

\abstract
{Angular momentum is a fundamental property that shapes the evolution of disc galaxies, strongly influencing the internal mechanisms that regulate star formation. Its content within disc galaxies is predicted to change over time, mainly as a result of external processes that regulate galaxy evolution. While several numerical studies paint a complex picture of angular momentum variation with environmental mechanisms, a recent observational finding suggests that galaxies are subject to angular momentum loss when they undergo interactions.}
{By studying the stellar angular momentum of simulated disc galaxies selected at various degrees of isolation, we aim to investigate whether isolation affects the stellar angular momentum content of disc galaxies and assess whether the environmental trends previously reported for baryonic angular momentum may also be reflected exclusively in the stellar component.}
{We selected star-forming disc galaxies in the IllustrisTNG simulation suite, for which we computed an isolation parameter based on local density. Using a density threshold, we identified isolated discs from non-isolated galaxies and performed a comparative study of their angular momentum content against other evolutionary parameters.}
{We find that isolation alone does not define the angular momentum content of a galaxy. Rather, whether a disc is gas-rich or gas-poor is directly linked to the specific angular momentum content, \js, of its stellar disc.}
{}


\maketitle
%
%-------------------------------------------------------------------

\section{Introduction}\label{sec:intro}

The evolution of galaxies is dictated by the interplay between internal and external processes, both contributing to shaping their morphologies and star formation properties. Because the nature and magnitude of external processes mainly depend on the environment, their effects on galaxies are significantly impacted by the local number density and gas conditions of their immediate vicinity \citep[e.g.][]{Dressler1980,Cayatte1990,Goto2003}. These effects can determine how the most basic properties of galaxies, such as baryonic mass, evolve over time. Similarly to mass, angular momentum ($j$), as a fundamental property, enables the characterisation of galaxies. This is because its quantity within a galaxy contributes to shaping its evolutionary path \citep[e.g.][]{Hernandez2006,Obreschkow2014}. In an ideal closed system, the total angular momentum is conserved; however, since galaxies interact with their environments, their angular momentum is expected to vary throughout their lifetime. Angular momentum is contained in both the baryonic and dark matter (DM) components of galaxies. In particular, baryonic angular momentum is mainly carried by stellar and gas discs.

Despite numerous observational and numerical studies, the existence of a potential correlation between the angular momentum and the environment of galaxies is unclear. For example, a few studies performed on galaxies of diverse morphologies suggested that the environment only weakly affects angular momentum (or the spin parameter), claiming that fundamental properties such as stellar mass and age are rather the primary drivers of galaxy spin \citep{Veale2017,Croom2024}. On the other hand, gas accretion events are believed to increase the angular momentum of galaxies when they acquire higher angular momentum gas from the circumgalactic medium (CGM) during their interactions with the environment, especially in cosmic-web filaments \citep[e.g.][]{Danovich2015,Stewart2013,Stewart2017}. In high-density regions, galaxies are subject to gravitational interactions with their neighbours, often resulting in galaxy-galaxy mergers. Studies performed on cosmological simulations show that, under certain conditions, galaxy mergers can effectively reduce the specific angular momentum of stars, $j_{\rm star}$ \citep{Lagos2017,Lagos2018}. Furthermore, our recent observational study \citep[][hereafter S24]{Sorgho2024} tentatively demonstrates that isolated discs free of major mergers over the past ${\sim}3$~Gyr, possess a higher baryonic angular momentum, $j_{\rm bar}$, than their non-isolated counterparts of the same baryonic mass. This implies that over their lifetimes, galaxies are subject to angular momentum loss due to interactions with other galaxies, consistent with the above cosmological simulations. However, as noted in \citet{Lagos2018}, several factors can affect this trend, ranging from the nature of the interactions to the orbital configurations of the interacting galaxies. Indeed, \citet{Lagos2018} find that galaxy mergers involving substantial amounts of gas can increase $j_{\rm star}$ when the spin vectors of the merging galaxies are aligned.

This picture of angular momentum gain and loss in galaxies is far from simplistic and depends on several parameters. This study represents a follow-up of \citetalias{Sorgho2024}, in which we investigate variations in the specific angular momentum content of stellar discs. In particular, through comparisons between observations and numerical simulations, we aim to investigate whether the \js\ content of isolated discs exhibits systematic differences compared to that of their non-isolated counterparts. Indeed, \citetalias{Sorgho2024} find that differences in the $j$ value between isolated and non-isolated galaxies are only observed for the overall baryonic components, not for stellar discs alone. In particular, we selected isolated galaxies from both the Analysis of the interstellar Medium of Isolated GAlaxies project \citep[AMIGA;][]{Verdes2005a} and the IllustrisTNG simulation suite \citep[hereafter TNG;][]{Nelson2018,Pillepich2018} as benchmarks to quantify potential effects of the environment on the angular momentum of stellar discs. AMIGA is a sample of highly isolated galaxies observed in various wavelengths, ranging from the optical to radio continuum regimes.

This paper is organised as follows. In \Cref{sec:data}, we describe the AMIGA sample together with the selection process for isolated subhaloes in the TNG simulation. In \Cref{sec:res}, we present a comparative analysis of the \js\ variations in both the observed and simulated galaxy samples. We discuss these variations in the context of galaxy evolution in \Cref{sec:disc} and summarise our findings in \Cref{sec:summary}.

%--------------------------------------------------------------------

\section{Data}\label{sec:data}

\subsection{TNG100 star-forming galaxies}\label{sec:data:tng}
The TNG Project \citep{Nelson2018,Pillepich2018} is a suite of gravomagneto-hydrodynamical simulations, a refined version of the earlier Illustris suite \citep{Vogelsberger2014,Genel2014}. It uses the moving-mesh code {\sc Arepo} \citep{Springel2010} and implements models for key physical processes relevant for galaxy formation and evolution \citep[see][for a description of the model]{Pillepich2018}. The project includes three simulation volumes: TNG50 (the highest resolution), TNG100 (intermediate resolution), and TNG300 (lowest resolution). In this work, since we are interested in searching for isolated galaxies in the simulation box, we used the TNG100 volume. Although this set does not offer the highest resolution, it represents a volume large enough to perform a robust search while offering a trade-off between resolution and box size. It consists of $2\times1820^3$ resolution elements in a ${\sim}100$~Mpc comoving box, with a mass resolution of $1.4\e{6}\,\Mo$. 

\begin{figure}
    \centering
    \includegraphics[width=\columnwidth]{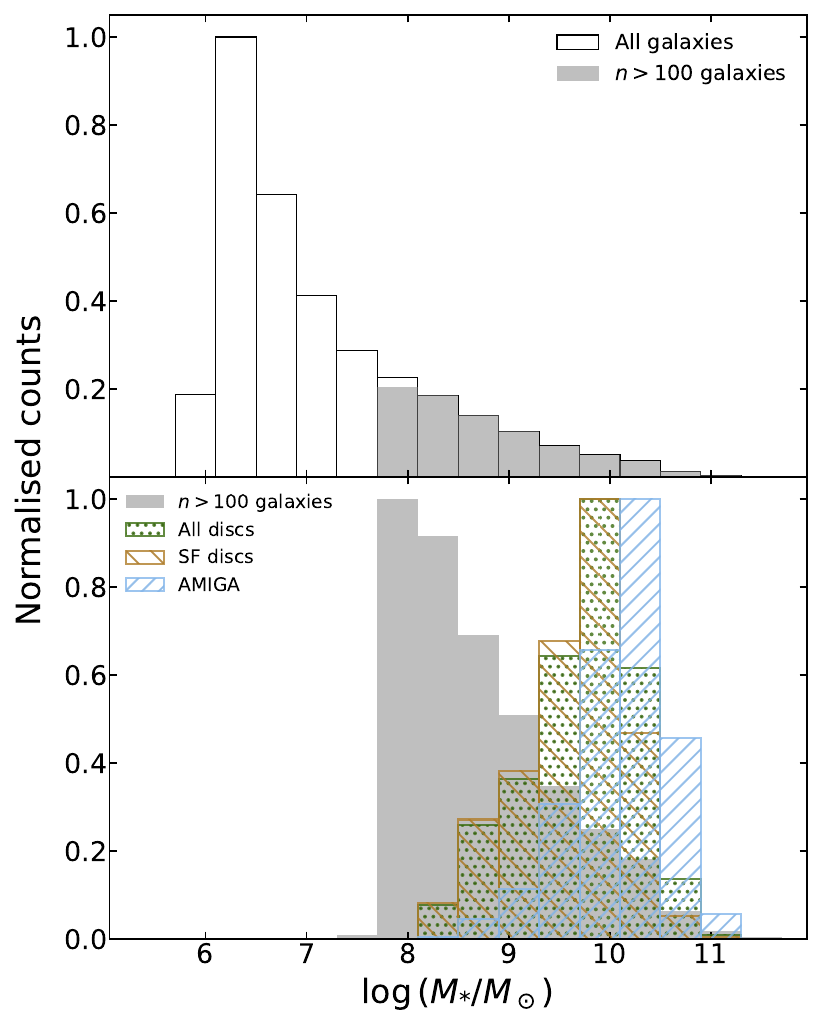}
    \vspace{-20pt}
    \caption{Top panel: Stellar mass distribution in the TNG100 simulation for all (empty histograms) and resolved subhaloes (grey histograms). Bottom panel: Same as in the top panel but for resolved galaxies (grey), all discs (green dots), and star-forming discs (orange hatches). The mass distribution of the AMIGA galaxies is also shown with blue-hatched histograms for comparison.}
    \label{fig:dist-sfd}
\end{figure}

The full TNG100 simulation contains ${\sim}4.4\e{6}$ confirmed subhaloes (i.e. galaxies) at redshift $z=0$. Of these, 49153 galaxies are resolved with a reasonable number (${\geq}100$) of stellar particles. We refer to this sample as ResTNG. To select star-forming disc galaxies, we proceeded as follows. 
First, we selected disc galaxies based on the prominence of the disc. This was estimated from the fractional mass of stars within the disc, whose circularity parameter is $\varepsilon>0.7$. The $\varepsilon$ parameter is defined as the ratio of the specific angular momentum of a star projected along the angular momentum axis of the galaxy to the maximum angular momentum of stellar particles with similar binding energies. This maximum was computed from the local neighbourhood of the star in a list sorted by binding energy, such that $\varepsilon = J_z/J(E)$ \citep{Lokas2020}. We set a threshold fractional mass of $f_{\varepsilon>0.7}>0.4$ for a galaxy to be considered disc-like. These values were calculated in \citet{Genel2015}, and a total of 4140 galaxies were found to meet these criteria. Next, we required that a disc galaxy have a specific star formation rate, $\rm sSFR \geq 10^{-11}\,yr^{-1}$, to be classified as star-forming. Here, we used the star formation rate (SFR) values computed from stars selected within twice the half-mass radius and averaged across the last 1~Gyr \citep{Donnari2019,Pillepich2019}. The sSFR threshold, commonly used in similar work, roughly corresponds to the minimum of the sSFR bimodal distribution of observed galaxies \citep[see][]{Kauffmann2004,Donnari2019,Walters2021}. These criteria reduce the sample size by a factor of ten, bringing the number of galaxies down to 3722. In other words, only ${\sim}10\%$ of resolved discs are non-star-forming. We denote this sample as SFTNG. In the top panel of \Cref{fig:dist-sfd}, we show how the stellar mass distribution of the star-forming discs compares to the overall TNG100 sample, highlighting subhaloes containing at least 100 stellar particles.
%--------------------------------------------------------------------

\subsection{Isolated galaxies in AMIGA and TNG100}\label{sec:data:isol}
The AMIGA sample \citep{Verdes2005a} was drawn from the Catalogue of Isolated Galaxies \citep[CIG;][]{Karachentseva1973}, which itself was constructed using stringent selection criteria. From an initial sample of 950 galaxies selected from 1050 objects, the sample was subsequently refined to include only galaxies satisfying strict isolation criteria and assessed using two principal parameters -- the local number density and the tidal forces exerted by neighbouring galaxies \citep{Sulentic2006,Verley2007a,Argudo-Fernandez2013}. Overwhelming evidence suggests that AMIGA galaxies constitute an almost nurture-free sample, exhibiting minimal values for parameters that are usually enhanced by interactions \citep[e.g.][]{Leon2008,Sabater2008,Lisenfeld2011,Jones2018}. Namely, the galaxies constituting the sample have undergone little environmental interaction in the last few gigayears. We began selecting our observed isolated galaxies from the AMIGA sample included in \citet{Jones2018}. The authors measured the \hi\ content of isolated galaxies, from which they constructed scaling laws predicting the atomic gas content of non-interacting galaxies. Of the 844 galaxies in the sample, we selected 599 with available WISE \citep[Wide-Field Infrared Survey Explorer][]{Wright2010} mid-infrared photometry. These photometric measurements enabled an estimation of the global stellar masses, as described in \citetalias{Sorgho2024}. Next, we discarded 151 galaxies with low (${<}30^\circ$) and high (${>}85^\circ$) inclination angles to avoid large uncertainties in inclination correction. The remaining 448 galaxies constitute our final sample of isolated galaxies, whose stellar mass distribution is shown in the bottom panel of \Cref{fig:dist-sfd} along with those of TNG100's star-forming discs.

\begin{figure}
    \centering
    \includegraphics[width=\linewidth]{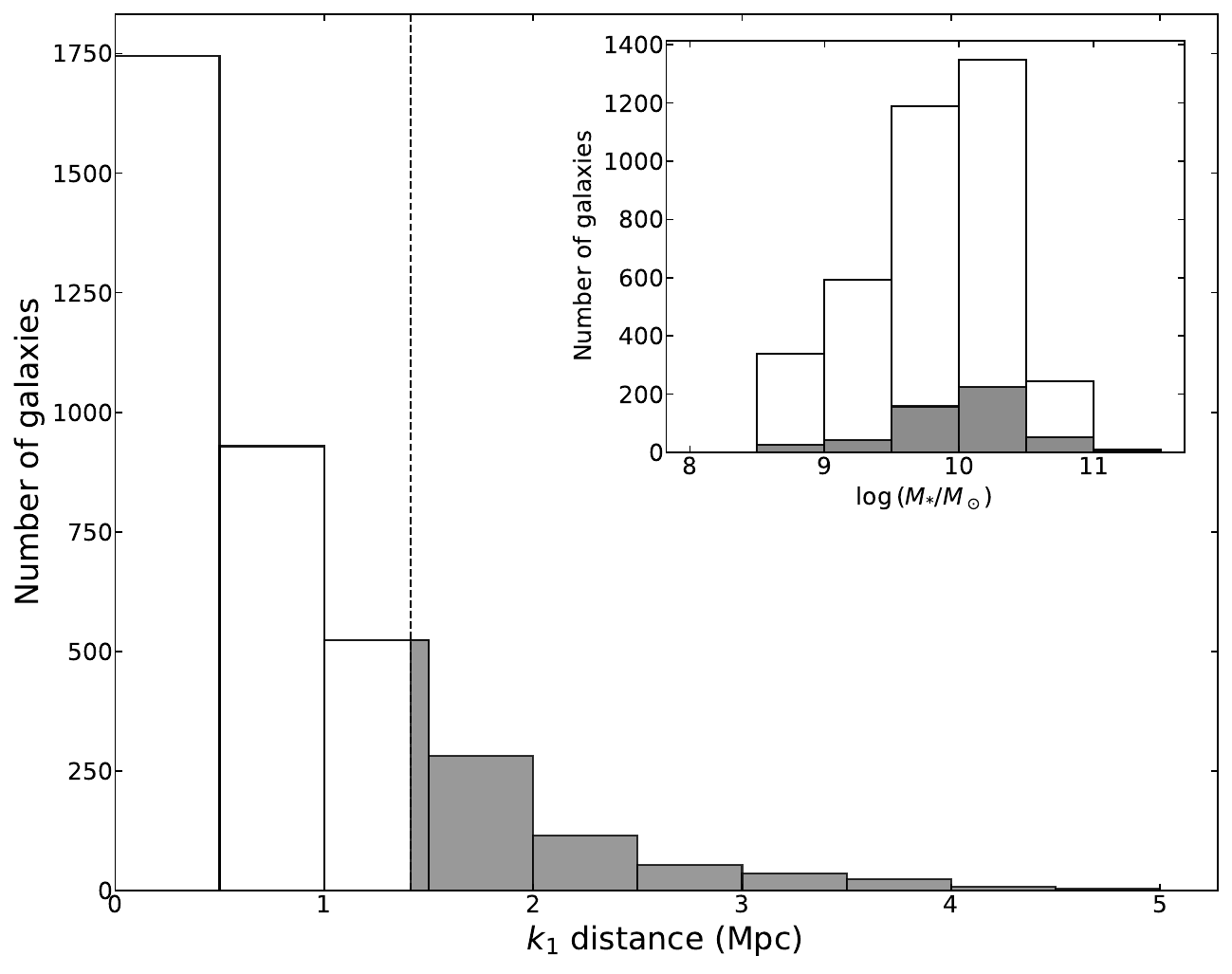}
    \vspace{-20pt}
    \caption{Distribution of the distance to the nearest neighbour for SF galaxies in the TNG100 simulation. The vertical dashed line shows the 84th percentile distance $k_1=1.4$~Mpc, and the grey area of the histogram highlights the fraction of galaxies with no neighbours within that distance. Inset: Stellar mass distribution of the TNG100 SF discs (empty histogram) and TNG100 isolated sample (grey histogram).}
    \label{fig:kdist}
\end{figure}

For the simulated TNG isolated galaxies, we adopted a method that closely models the observed AMIGA galaxies. First, we defined central galaxies with no neighbours within a defined volume as isolated. This is a stricter definition than that of \citet{Walters2021}, who considered central galaxies whose secondary members have less than ${\sim}20\%$ of their mass. To determine the radius of the volume within which neighbours are searched, we proceeded as follows. We calculated the 3D distance between each star-forming (SF) disc galaxy and its closest resolved (i.e. of more than 100 stellar particles) neighbour in the entire TNG100 sample. We find that ${>}70\%$ of SF discs have at least one neighbour within 1~Mpc and that the nearest distance distribution has an 84th percentile of $k_1=1.41\rm\,Mpc$ (see \Cref{fig:kdist}). In other words, only 16\% of the SF discs have no neighbour within a radius of 1.4~Mpc, equivalent to a standard deviation above the average distance between two closest neighbours. We note that these represent ${\sim}595$ galaxies that are, by definition, ${\sim}34\%$ more isolated than the average galaxy in the sample. To restrict the study to further isolated subhaloes, we set the minimum radius to $r_v = 2\,k_1$. That is, we considered a galaxy strictly isolated if it has no neighbour within a radius of 2.8~Mpc. We find that 99 galaxies satisfy this criterion. However, as noted in \citet{Karachentseva1973} and consistent with the AMIGA criteria \citep{Verley2007b,Verley2007a}, a galaxy can only be considered a major neighbour if its mass is at least one fourth of the galaxy of interest. Considering this, we expanded the sample to include galaxies whose neighbours within a radius of 2.8~Mpc have stellar masses below 25\% of their own mass. A total of 411 additional galaxies fulfil this criterion, bringing the sample size to 510 SF galaxies. For the remainder of the paper, these are referred to as the IsoTNG sample. The shape of their $\Ms$ distribution is similar to that of their parent SFTNG sample (inset panel of \Cref{fig:kdist}). We emphasise that these galaxies represent the most isolated candidates within the TNG100 volume according to our adopted criteria, but they are not intended to reproduce the AMIGA sample exactly. A more robust way to mimic the AMIGA sample would require a reprojection of all subhaloes onto the plane of the sky and converting the third physical dimension into systemic velocities. We summarise the different samples considered in the study in \Cref{tab:samples}.

\begin{table*}
\centering
\footnotesize
\caption{Summary of the simulated and observed samples considered in this study.}\label{tab:samples}
\begin{tabular}{l l l c c} 
\hline 
\hline 
\rule{0pt}{10pt}
Parent sample & Sample name & Criteria & Size & $\rm\log{(M_*/M_\odot)}$ range \\ 
\hline \rule{0pt}{10pt}
\multirow{5}{*}{TNG100} & FullTNG & Full TNG sample & $4.36\times10^{6}$ & [5.62, 12.38] \\ 
 & ResTNG & Resolved TNG galaxies & 49153 & [7.78, 12.38] \\ 
 & SFTNG & Star-Forming TNG discs & 3722 & [8.52, 11.39] \\ 
 & IsoTNG & Isolated SF TNG discs & 510 & [8.56, 11.39] \\ 
 & NoIsoTNG & Non-isolated SF TNG discs & 3212 & [8.52, 11.36] \\ 
 & TFTNG & bTFr sample & 409 & [8.56, 11.26] \\ 
\noalign{\vskip 5pt}
\multirow{2}{*}{AMIGA} & Main & Inclination-selected AMIGA discs & 448 & [8.55, 11.34] \\ 
 & S24 & AMIGA discs with resolved $\textsc{Hi}$ maps & 31 & [8.55, 11.34] \\ 
\hline
\end{tabular}
\end{table*}
%--------------------------------------------------------------------

\section{Variation of angular momentum}\label{sec:res}

\subsection{Atomic gas fractions: TNG100 versus observations}\label{sec:res:gasdist}
\begin{figure}
    \centering
    \includegraphics[width=\linewidth]{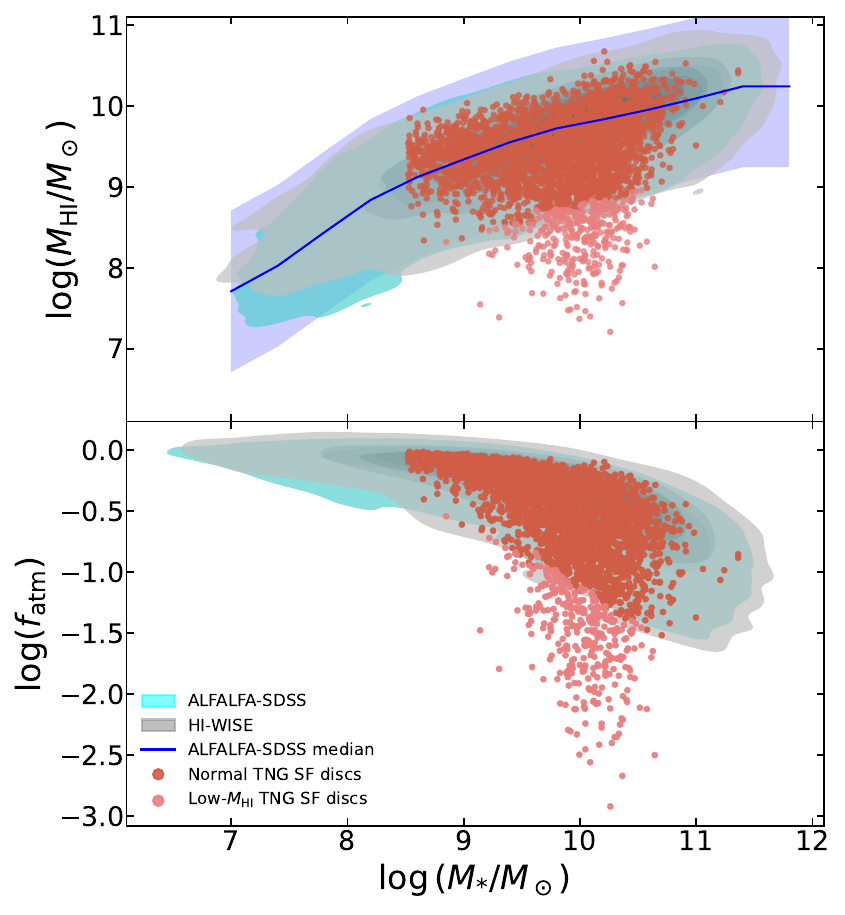}
    \vspace{-20pt}
    \caption{Variation of \hi\ mass (top panel) and atomic gas-to-baryonic mass fraction (bottom panel) as a function of the stellar mass for the simulated and observed galaxies.}
    \label{fig:hi-vs-star}
\end{figure}

The 3722 star-forming galaxies of the SFTNG sample exhibit a wide range of SF activity and varying \hi\ content. Their \hi\ masses, $\Mhi$, and total atomic gas to total baryonic fractions, $\Mhi/M_{\rm bar}$, are given in \Cref{fig:hi-vs-star}. For comparison, we overlay galaxies from the ALFALFA-SDSS \citep{Durbala2008} and \hi-WISE \citep{Parkash2018} catalogues. ALFALFA \citep[Arecibo Legacy Fast ALFA;][]{Giovanelli2005} is a blind \hi\ survey covering ${\sim}7000\rm\,deg^2$ of the northern sky and redshifts of up to 18,000~\kms. Thus, it provides an \hi-selected view of the nearby galaxy population. The ALFALFA-SDSS catalogue was constructed by searching through the Sloan Digital Sky Survey (SDSS) database for optical counterparts to the 31,501 \hi\ sources in the complete ALFALFA sample \citep{Haynes2011,Haynes2018}. \citet{Durbala2008} derived the optical properties of the catalogue based on SDSS, \wise\, and GALEX photometries. They found that 28,267 galaxies had a clear SDSS counterpart for which the optical photometry had the highest quality. The remaining galaxies either had large photometric uncertainties, dubious SDSS counterparts, or fell outside the SDSS footprint. For this work, we used the sample of galaxies with clear optical counterparts. Furthermore, for consistency with the AMIGA sample, whose stellar masses were estimated from \wise\ photometry, we used the author's $\Ms$ estimates based on \wise's W1 band and employing the prescription of \citet{McGaugh2005}.

Similarly to ALFALFA-SDSS, the \hi-\wise\ catalogue is based on the untargeted \hi\ survey of HIPASS \citep{Meyer2004}. It contains 4,135 \hi\ sources matched against several optical catalogues for stellar counterparts, for which the \wise\ photometry was measured. The final sample size, after imposing declination and signal-to-noise ratio cuts, includes 3,158 galaxies. The stellar masses of these galaxies were also estimated from the \wise\ photometry. In \Cref{fig:hi-vs-star}, we show how the simulated TNG galaxies compare with the two \hi-selected samples in terms of \hi\ mass and atomic gas fraction. The distributions of the \hi\ contents of both the ALFALFA-SDSS and \hi-\wise\ samples are broadly similar, showing a gradual increase from low $\Ms$ to more massive galaxies. Most of the galaxies in these samples lie within 1~dex of the ALFALFA-SDSS's median at all stellar mass bins. Although the SF discs of the TNG100 sample shows a similar trend, it does contain a number of galaxies forming a vertical tail at $\Ms{\sim}10^{10}\,\Mo$, with \hi\ masses as low as ${\sim}10^7\,\Mo$. These galaxies likely possess \hi\ masses significantly lower than what their stellar masses allow. By considering the ALFALFA-SDSS sample as a reference for \hi\ content, we identified these low-$\Mhi$ by setting a cut at 1 dex below the median $\Mhi$ at a given mass bin. That is, any galaxy with $\Mhi$ less than 10\% of that expected from the ALFALFA-SDSS trend was considered an extremely gas-poor galaxy. Of the 3722 SFTNG galaxies, 430 obey this condition. These galaxies have atomic gas fractions ranging from 0.0 to ${\sim}0.29$.

%--------------------------------

\subsection{Angular momentum measurements}\label{sec:res:angmom}
\begin{figure}
    \centering
    \includegraphics[width=\linewidth]{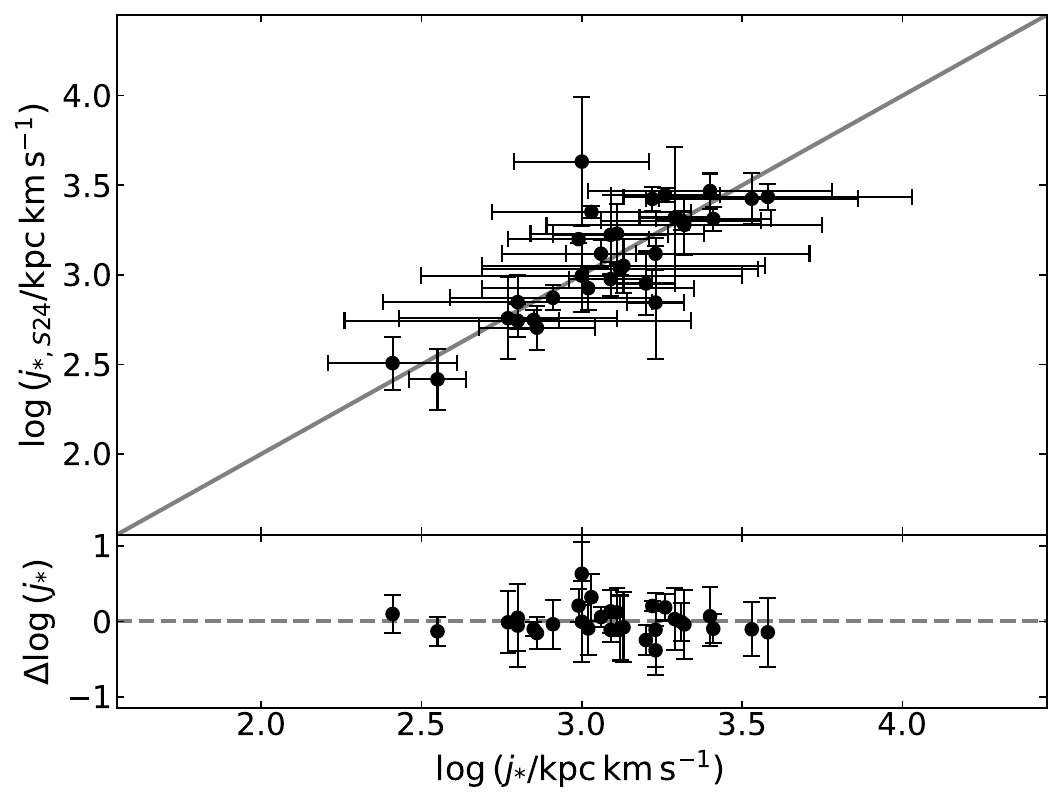}
    \vspace{-20pt}
    \caption{Integrated specific stellar angular momentum \citepalias[$j_{\rm *,S24}$;][]{Sorgho2024} vs the analytical \js\ (top panel) and their residual (bottom panel).}
    \label{fig:js-comp}
\end{figure}

For a galaxy rotating constantly at a velocity $V_{\rm circ}$ and scale radius $R_\star$, the specific angular momentum of its stellar component can be computed using the approximation \citep{Fall1983,Mo1998,Romanowsky2012,Obreschkow2014},
\begin{equation}\label{eq:jobs}
    j_\star = 2\,R_\star\,V_{\rm circ}.
\end{equation}
We refer to these quantities as analytical $j_\star$ values. As noted in \citet{Obreschkow2014}, the scale radius can be approximated as $R_\star \simeq 0.3\,R_{25}$, where $R_{25}$ is the $B$-band radius at the $25\,\rm mag\,arcsec^{-2}$ isophote. The circular velocity is taken as half of $W_{50}$, the velocity width measured at the 50\% flux level. The velocity widths of AMIGA galaxies were derived in \citet{Jones2018} from single-dish \hi\ profiles. Of the 515 galaxies of the AMIGA sample, 31 galaxies for which spatially resolved \hi\ and optical maps are available, are included in \citetalias{Sorgho2024}. Integrated, robust specific angular momenta of these galaxies were measured therein. To test the robustness of the analytical results, we provide a comparison with the values measured by \citetalias{Sorgho2024} in \Cref{fig:js-comp}.

\begin{figure}
    \centering
    \includegraphics[width=\linewidth]{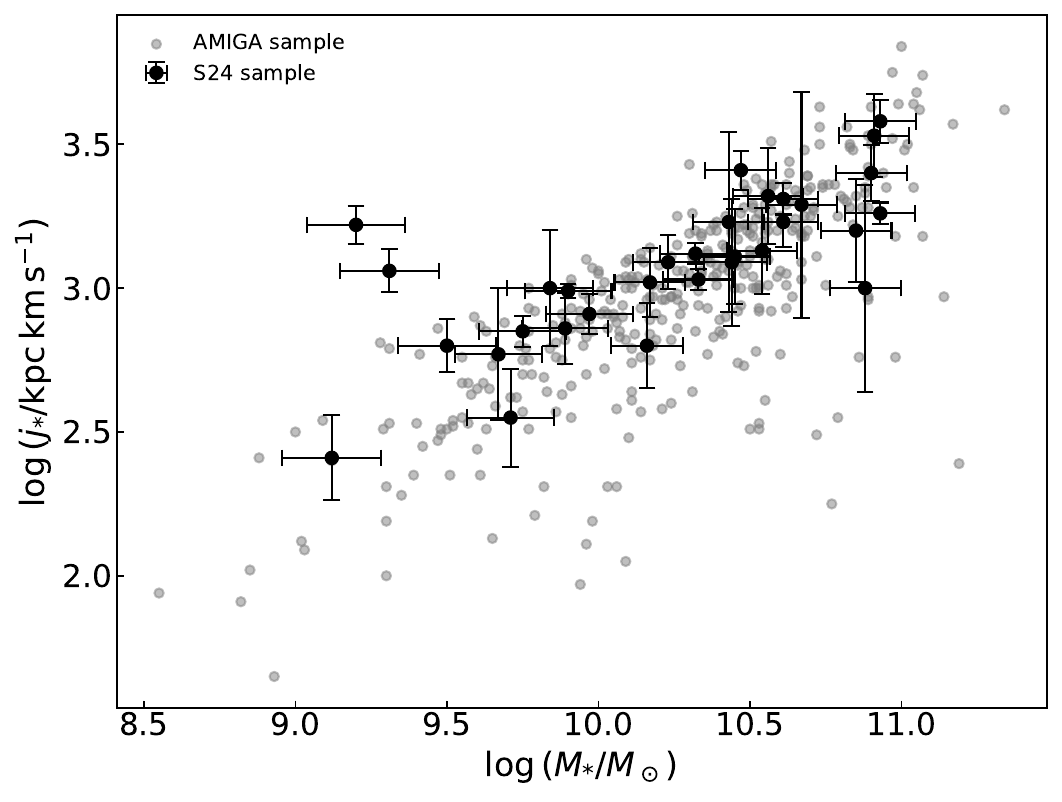}
    \vspace{-20pt}
    \caption{Specific angular momentum as a function of stellar mass for the AMIGA galaxies. The galaxies included in \citetalias{Sorgho2024} (integrated \js), highlighted in the plot, are representative of the sample in terms of mass and angular momentum distribution.}
    \label{fig:jvsm-amiga}
\end{figure}

Furthermore, as depicted in their distributions on the $M_{\rm*}$-$j_{\rm*}$ plane (\Cref{fig:jvsm-amiga}), the \citetalias{Sorgho2024} sample is representative of the AMIGA galaxies included in the present study. They both span similar stellar mass ranges, and most of the \citetalias{Sorgho2024} sample galaxies are located on the main plane formed by the wider sample considered in the present work.

The specific angular momenta of the TNG100 galaxies were taken from the TNG's value-added catalogues. They were measured from gravitationally bound stellar particles contained within ten times the effective radius of the subhalo \citep{Genel2015}. In a given subhalo, the particle with the lowest gravitational potential was considered the galactic centre and thus the reference for the calculation. The specific stellar angular momenta were then calculated as \citep[e.g.][]{DeFelippis2020}
\begin{equation}
    j_{*} = \frac{1}{M_{*}} \sum_{i=1}^{N} m_i \left( \mathbf{r}_i - \mathbf{r}_0 \right) \times \left( \mathbf{v}_i - \mathbf{v}_{\mathrm{cm}} \right),
    \label{eq:jtng}
\end{equation}
where $m_i$, $\mathbf{r}_i$, and $\mathbf{v}_i$ are the mass, position, and velocity, respectively, of particle $i$. Here, $\mathbf{r}_0$ and $\mathbf{v}_{\mathrm{cm}}$ are the central position and centre-of-mass velocity, respectively, of the subhalo. As noted in \Cref{sec:data:tng}, the calculations were limited to resolved (${\geq}100$ stellar particles) galaxies with stellar masses $\Ms\geq3.4\e{8}\Mo$.
%--------------------------------

\subsection{TNG100 galaxies on the baryonic Tully-Fisher relation}\label{sec:res:btfr}
\begin{figure}
    \centering
    \includegraphics[width=\columnwidth]{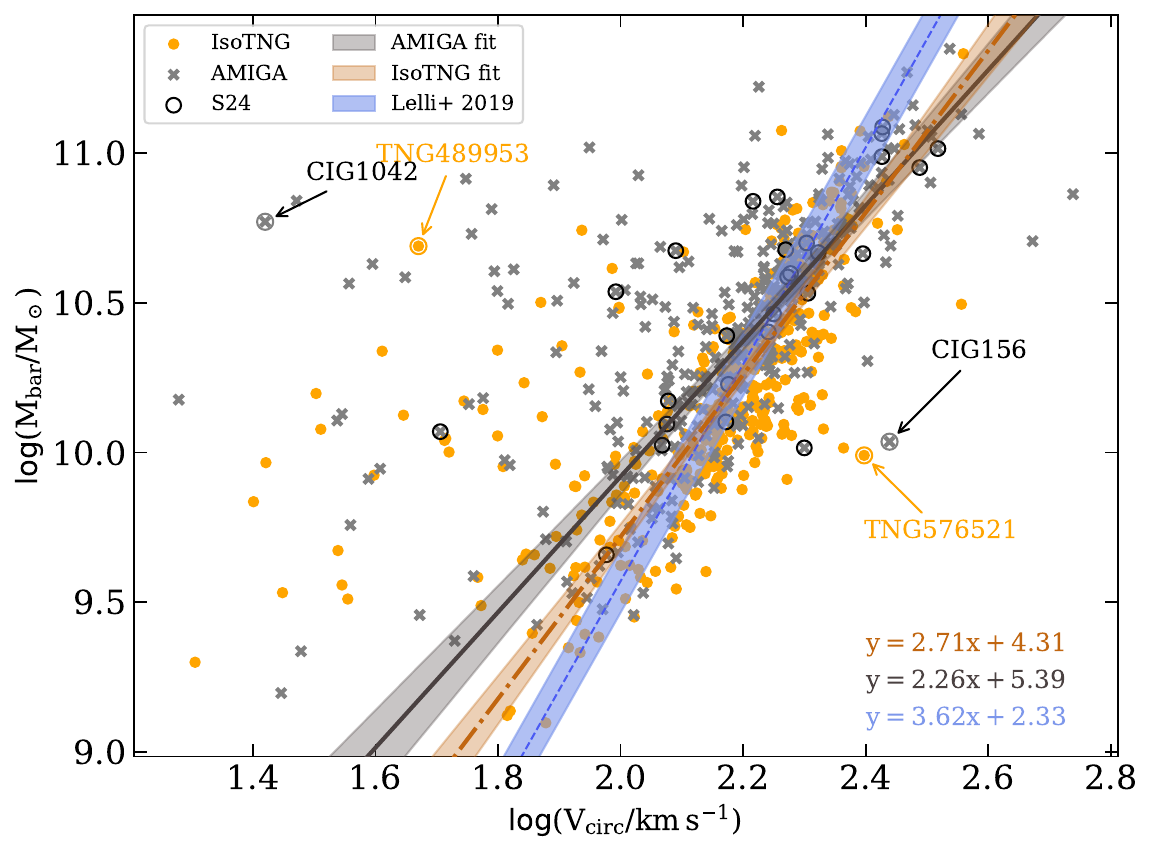}
    \vspace{-20pt}
    \caption{bTFr of the observed AMIGA and simulated TNG isolated galaxies. The relation of \citet{Lelli2019} is overlaid for comparison.}
    \label{fig:btfr}
\end{figure}

The difference in the measurements of \js\ between the simulated and observed galaxies can cause inconsistencies in their interpretations. To compare the kinematics of the two samples, we placed them on the baryonic Tully-Fisher relation (bTFr): the total baryonic mass as a function of the rotation velocity \citep{Tully1977,McGaugh2000,Bell2001,McGaugh2015}. Consistent with the definition of \js\ in \Cref{eq:jobs}, we define the rotation velocity as half the \hi\ linewidth. For the TNG100 galaxies, we followed the prescriptions of \citet{Baes2025} to derive $W_{50}$. We used the radiative transfer code {\sc SKIRT} \citep{Baes2011,Camps2015,Camps2020} to simulate, from an observer's position, the \hi\ profiles of the TNG100 subhaloes. As a general, multipurpose Monte Carlo tool, SKIRT is commonly used to generate synthetic data for subhaloes obtained from cosmological hydrodynamical simulations. It incorporates different stellar populations and the complex star–dust geometry and can be used to simulated multiwavelength views of galaxies based on radiative transfer models, including dust, X-ray, atomic, and molecular gas. We ran the simulation on all identified isolated galaxies in TNG100. For each we selected all the gravitationally bound gas cells. Since we are interested in the gas component of the galaxies, we ran SKIRT in the \verb+GasEmission+ mode. In this mode, \hi\ emits photon packets that propagate to the observer, accounting for Doppler shifts and thermal broadening \citep{Camps2020}. This results in a \hi\ line profile `seen' from the observer's perspective, from which we estimated $W_{50}$ at half the peak \hi\ flux. \citet{Baes2024} define five arbitrary observer positions (O1 to O5), each at a distance of 206.26~Mpc, regardless of the orientation of individual galaxies. Four of these positions are independent and were chosen to allow optimal arrangement, with the fifth being antipodal to one of them. In this work, we chose the single position O1 (polar angle of $60.23^\circ$ and  azimuth of $99.80^\circ$) for our analyses. In principle, the choice of the viewing position should not affect our remaining analysis, provided that the same position is consistently considered (see discussion in \Cref{sec:app:or}). Furthermore, to avoid large uncertainties on the inclination correction of $W_{50}$, we discarded galaxies with inclinations below $30^\circ$. Similarly, following \citet{Baes2025}, we also excluded edge-on galaxies at inclinations above $85^\circ$ to avoid large dust attenuation uncertainties. In total, 101 galaxies were excluded based on these inclination criteria, bringing the sample size down to 409. We note that this represents a TNG100 sample of isolated galaxies eligible only for the bTFr analysis (hereafter, TFTNG sample), but not for the rest of the paper. We show the relation for these in \Cref{fig:btfr}, overlaid on the 448 observed AMIGA galaxies obeying the same inclination criteria. Although the simulated and observed galaxies exhibit linear regressions of different slopes in the bTFr plane, their baryonic masses are broadly consistent within the ranges of typical velocity widths (up to $V_{\rm circ}{\sim}250$~\kms).

\begin{figure*}
    \centering
    \includegraphics[width=\textwidth]{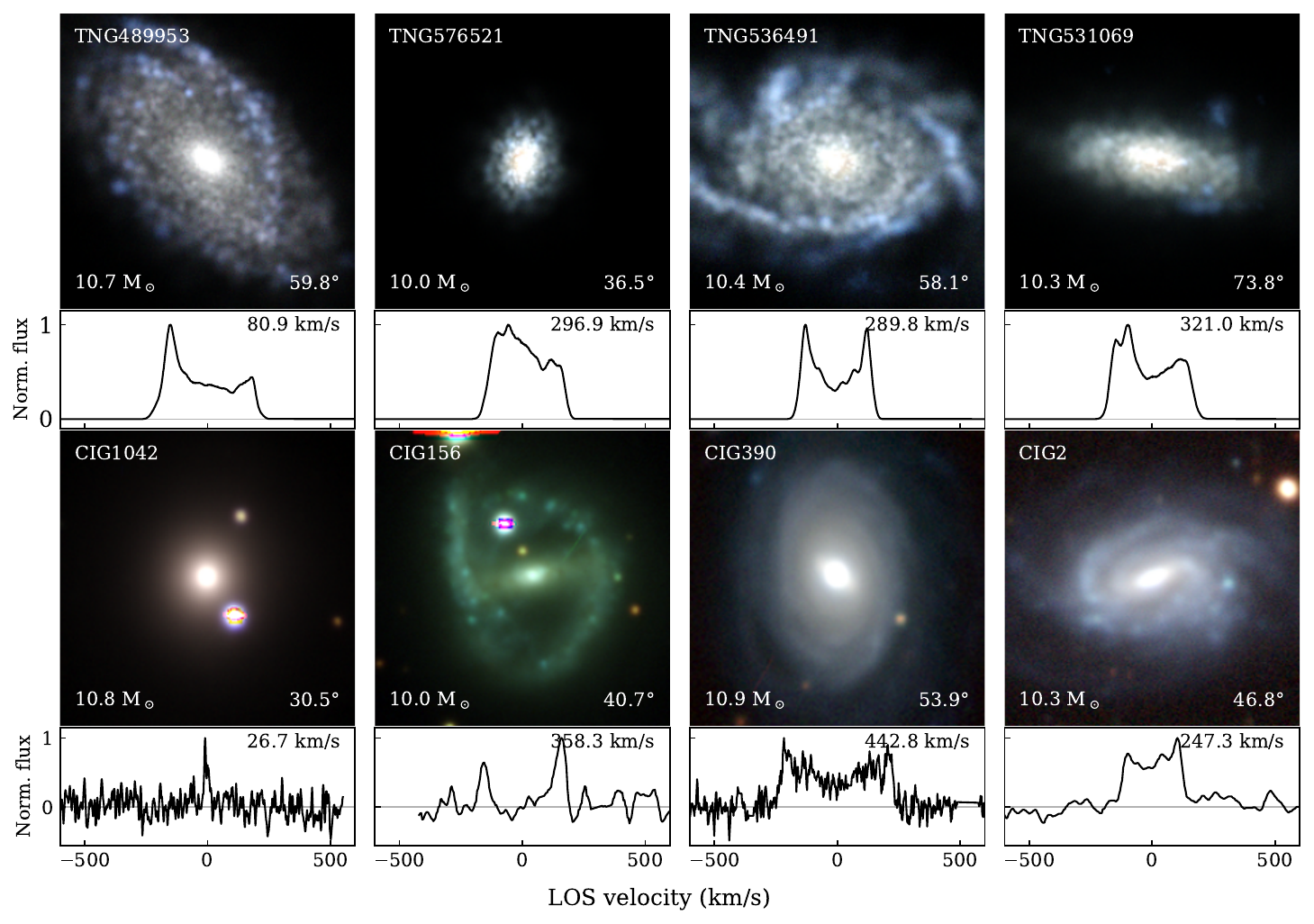}
    \vspace{-20pt}
    \caption{Top row: SKIRT synthesised composite g-, r-, and z-band images of four isolated TNG100 galaxies and their \hi\ profiles. The inclinations from the O1 observer's perspective are shown in the bottom-right corners of the images, and the inclination-corrected velocity widths in the top-right corners of the \hi\ profile panels. From left to right, the first two galaxies deviate most from the bTFr, while the last two conform to the relation. The respective TNG100 IDs are given above the synthesised images of the galaxies. Bottom row: Same as the top row for the observed AMIGA galaxies but with composite images obtained from the DECaLS grz bands.}
    \label{fig:skirt-profs}
\end{figure*}

Furthermore, the \citetalias{Sorgho2024} galaxy sample are spread throughout the parameter space and do not favour any particular regions of circular velocity or baryonic mass. In the figure, we highlight the two galaxies that deviate primarily from the linear regression line $\log{M_{\rm bar}} = \alpha\log{V_{\rm circ}} + \beta$ representing the bTFr trend of the TNG100 isolated galaxies. The synthesised optical images and \hi\ profiles of both galaxies, as well as those of two galaxies randomly selected on or close to the line of best fit, are shown in \Cref{fig:skirt-profs}. Although not immediately evident from its synthetic optical image, the global \hi\ profile of the TNG subhalo (ID 489953) exhibiting the largest deviation at the low-velocity end reveals a highly asymmetric gas distribution, which biases its velocity width measurement. Consequently, the $W_{50}$ parameter was determined from a single peak, yielding a value of 80.9~\kms. For comparison, the $W_{20}$ measurement of the subhalo encompasses both `horns' of the profile, resulting in a velocity width as large as 402.0~\kms. In contrast, its observed counterpart, CIG 1042, appears to be an early-type (S0) galaxy with a single-peaked global \hi\ profile. Despite its relatively high stellar mass ($\Ms=10^{10.8}\Mo$), its exceptionally narrow \hi\ line width places it as a clear outlier in the bTFr.

%--------------------------------------------------------------------

\subsection{Fall relation: AMIGA versus TNG}
\begin{figure*}
    \centering
    \includegraphics[width=\textwidth]{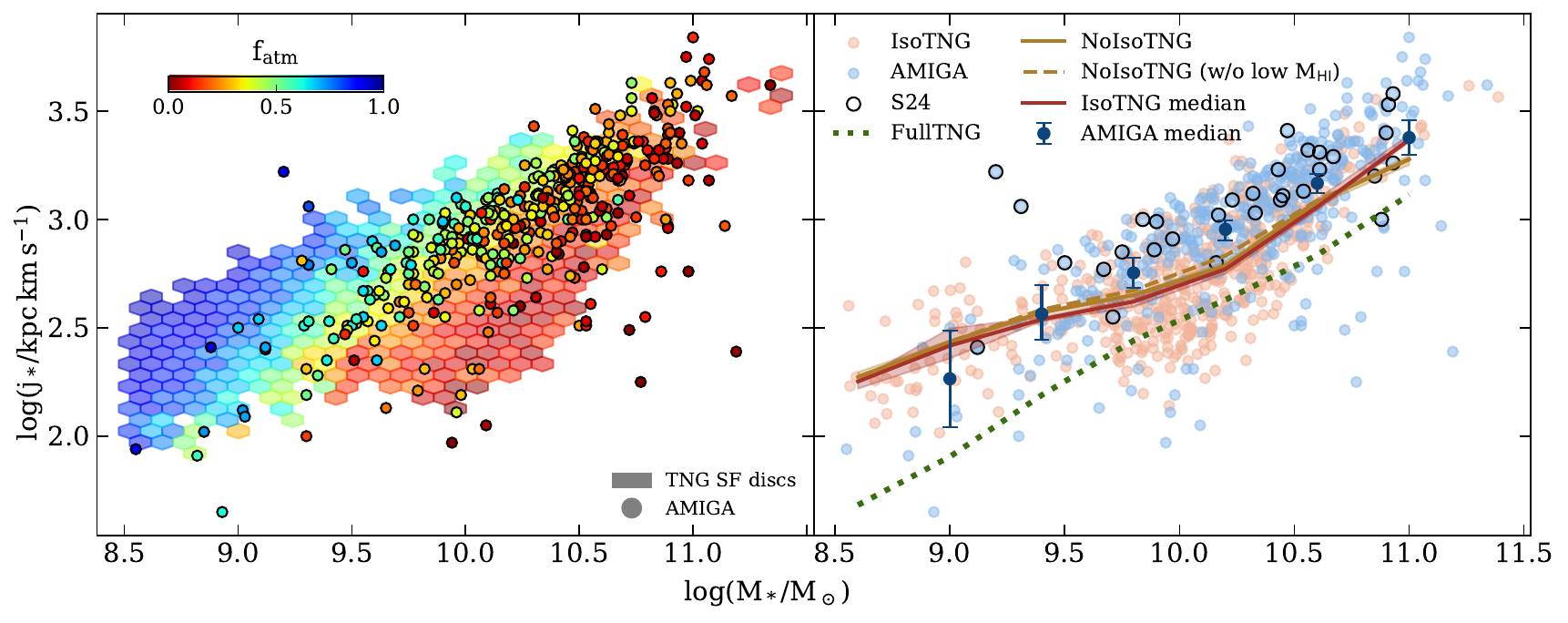}
    \vspace{-20pt}
    \caption{Specific angular momentum, \js\,, as a function of stellar mass for both simulated TNG100 and AMIGA discs. Left: Galaxies colour-coded with the atomic gas fraction $f_{\rm atm}$. Right: Running medians of different selections shown for comparison (see \Cref{sec:res:angmom} for details).}
    \label{fig:jvsm-tng-obs}
\end{figure*}

In the left panel of \Cref{fig:jvsm-tng-obs} we show the Fall relation---angular momentum as a function of the stellar mass---for the simulation and observational data, colour-coded with the atomic gas fraction (the ratio of atomic gas to total baryonic mass). The figure reveals a broad agreement between the trends of both the TNG and AMIGA samples and shows that, consistently with past studies (e.g. \citealp{ManceraPina2021a}, \citetalias{Sorgho2024}), gas-rich galaxies possess more angular momentum than their gas-poor counterparts of the same stellar mass. Furthermore, as expected, we note that disc galaxies have higher angular momenta than the overall sample, at any given stellar mass (right panel of \Cref{fig:jvsm-tng-obs}). Overall, the variation of the specific angular momentum content of the isolated discs is similar to that of non-isolated SF discs (NoIsoTNG sample). This contradicts the trends found for the baryonic angular momentum for observed galaxies \citepalias{Sorgho2024}, where isolated discs were found to conserve a higher fraction of their angular momentum than their non-isolated counterparts. Furthermore, at stellar masses of ${\sim}10^{10}\Mo$, both isolated and SF disc samples of TNG100 present a turnover region (a `dip') in their angular momentum content, suggesting a population of low-\js\ galaxies in that mass range. This turnover occurs at the same stellar mass range where the low-$\Mhi$ galaxies are observed in the $\Mhi{-}\Ms$ plane (\Cref{fig:hi-vs-star}). Furthermore, when low-$\Mhi$ objects are removed from the TNG100 SF discs, a slightly shallower but still existing turnover is observed in the median \js\ values.

%--------------------------------------------------------------------

\section{Discussion}\label{sec:disc}

\subsection{Angular momentum content as a function of isolation}
In a previous observational study on the environmental impacts on galaxy angular momentum, we found that isolated galaxies possess higher contents of baryonic angular momentum than non-isolated galaxies of the same baryonic mass \citepalias{Sorgho2024}. However, when the baryonic angular momentum is broken down into its individual components, these environmental trends disappear. These specific angular momenta of the AMIGA galaxies were measured out to the last available radius, typically beyond the 25th magnitude. In \Cref{fig:jvsm-tng-obs} (right panel), we note that the isolated galaxies in the TNG100 simulations do not exhibit systematic differences with respect to their non-isolated counterparts in terms of specific stellar angular momentum content. Rather, the median \js\ values of both isolated and non-isolated galaxies overlap throughout the entire stellar mass range. Similar to AMIGA galaxies, the \js\ values of the TNG100 subhaloes are expected to encapsulate the total angular momentum of the stellar component, since they were measured within volumes as large as 10 half-light radii. A Kolmogorov–Smirnov (KS) test applied to the cumulative distribution functions (CDFs) of the \js\ values for the isolated and non-isolated samples yields a low test statistic of $D{=}0.06$ with an associated $p$-value of $p{=}0.53$. We recall that $D$ quantifies the maximum absolute difference between the two CDFs over the entire range of \js, implying that the CDFs differ by no more than ${\sim}6.4\%$ at any given \js\ value. Furthermore, the probability ($p$) of obtaining a test statistic $D$ at least as large as the observed value, under the null hypothesis that both samples are drawn from the same parent distribution, is 53\%. Since this exceeds the conventional significance threshold of 5\%, we cannot reject the null hypothesis, and therefore the two CDFs are statistically consistent. Consequently, any differences between the \js\ distributions of isolated and non-isolated galaxies are not statistically significant.

\begin{figure*}
    \centering
    \includegraphics[width=\textwidth]{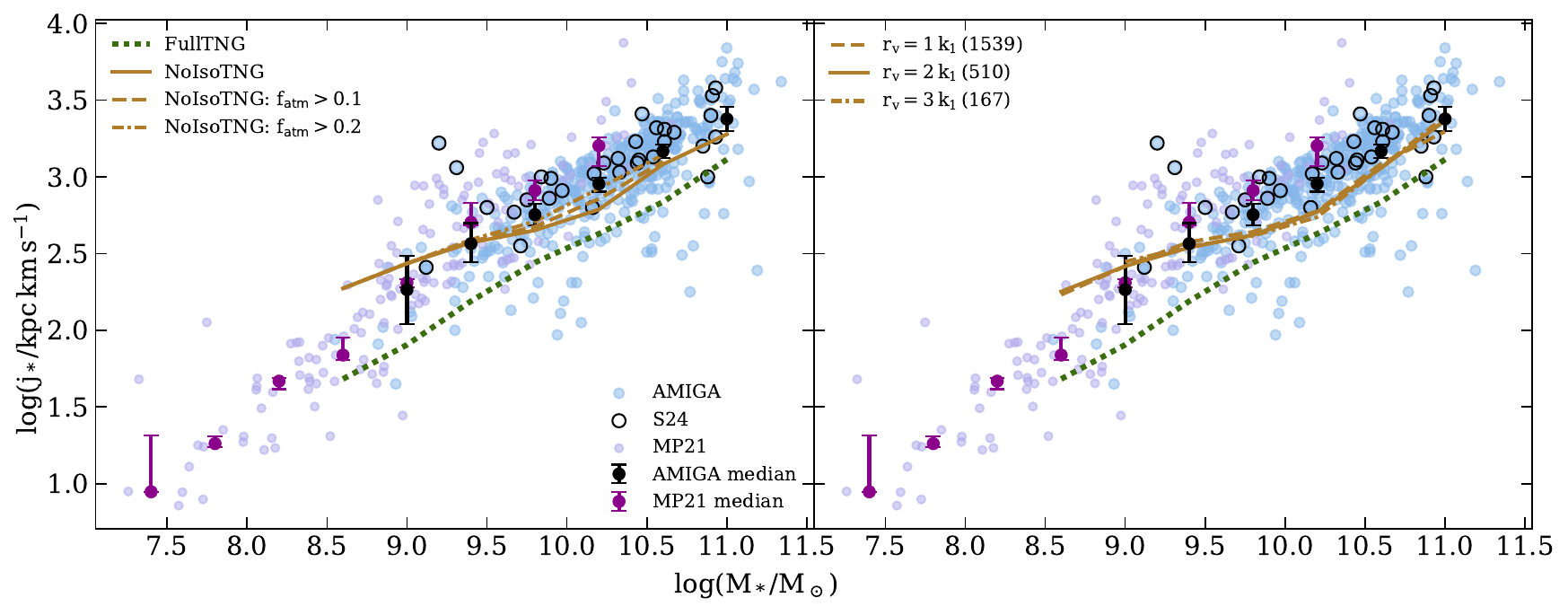}
    \vspace{-20pt}
    \caption{Same as the right panel of \Cref{fig:jvsm-tng-obs} but with different thresholds of $f_{\rm atm}$ (left panel) and various radial criteria for isolation (right panel). In addition, the sample of local galaxies from \citet{ManceraPina2021} is included for comparison.}
    \label{fig:jvsm-tng-fatm}
\end{figure*}

Furthermore, we find that most AMIGA galaxies lie above the median TNG SF discs in the \js--$\Ms$ parameter space. The differences are most prominent for TNG galaxies with stellar masses of ${\sim}10^{10}\Mo$, where a dip is observed. The removal of extremely low-\hi-mass galaxies from the TNG sample does not significantly affect the median \js\ trend at the stellar mass range of the dip, hinting at the presence of low-\js\ galaxies with typical \hi\ masses in the simulated sample.
To further investigate the cause of the turnover region, we excluded gas-poor subhaloes from TNG100 based on threshold $f_{\rm atm}$ values. The left panel of \Cref{fig:jvsm-tng-fatm} shows that TNG100 only overlaps with AMIGA when $f_{\rm atm}\gtrsim 0.2$. However, of the 225 simulated galaxies with atomic gas fractions higher than the threshold, only 13 have stellar masses $\Ms\gtrsim10^{10.5}\,\Mo$. The sample is predominantly composed of moderately sized galaxies. Finally, the figure shows that the \js\ values of the AMIGA sample are consistent with those of non-isolated galaxies from the literature \citep{ManceraPina2021} in the stellar mass regions where the two populations overlap.

\begin{figure}
    \centering
    \includegraphics[width=\columnwidth]{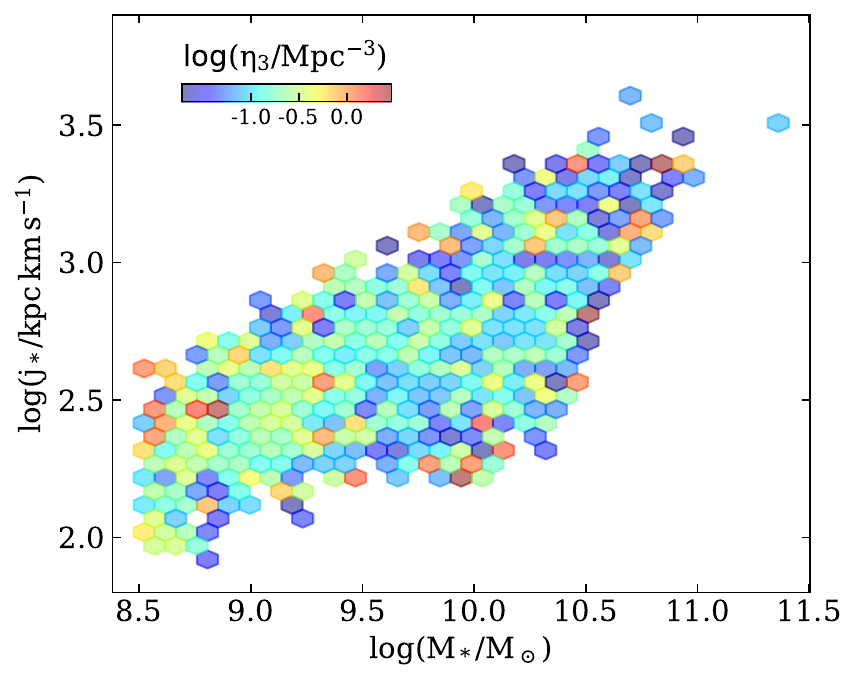}
    \vspace{-20pt}
    \caption{Same as the left panel of \Cref{fig:jvsm-tng-obs} but colour-coded with the local density number, $\eta_3$.}
    \label{fig:mj-n}
\end{figure}

In our previous study, we demonstrated that isolated galaxies selected from the AMIGA sample exhibit higher baryonic angular momenta than their non-isolated counterparts of the same mass \citepalias{Sorgho2024}. In this work, we find that the \js\ content of isolated TNG galaxies is not systematically higher than those of non-isolated subhaloes. While these results do not directly contradict earlier findings, they suggest that isolation is not the main external factor regulating angular momentum in galaxies. To further verify this, we considered the nearest-neighbour density metric $\eta_N=N/V_N$, where $V_N=4\pi D_N^3/3$ is the volume enclosing the nearest $N$ neighbours. We emphasise that for a given galaxy of (optical) diameter $d$, only neighbours with sizes ${\geq}0.25d$ were considered. Evidently, a low number $N$ probes the immediate neighbour, while a large $N\ ({\gtrsim}5)$ surveys a considerable volume, comparable to or approaching the large-scale structure. To get an estimate of the local density, we considered the three nearest neighbours ($N{=}3$). This is large enough to enable a distinction between field and group galaxies \citep[e.g. Hickson Compact Groups have four to ten members;][]{Hickson1982} but smaller than the scale of the large-scale environment. In \Cref{fig:mj-n}, we reproduce the Fall relation in \Cref{fig:jvsm-tng-obs} but colour-coded with $\eta_3$. Consistent with the above result, no unambiguous correlation is observed in the \js\ distribution. Low-density (low $\eta_3$) subhaloes are mostly found at the high-$\Ms$ end of the parameter space, but no overall variation of \js\ with $\eta_3$ is observed. This is further supported by the weak Spearman's rank correlation coefficient of $\rho=-0.08$ ($p=1.8\times10^{-5}$) between the two variables.

%---------------------------

\subsection{Dependence of \js\ variation on the isolation definition}
As noted in \Cref{sec:data:isol}, the definition of isolation for TNG100 galaxies does not correspond to that used in the AMIGA selection. The 3D radius within which neighbours are searched is based on the galaxy distribution in the volume. It is hence statistical rather than based on a physically motivated distance threshold. That is, we are interested in the most isolated galaxies of the sample, which do not necessarily correspond to the observed AMIGA galaxies. The 84th percentile of the distribution of the separation distance between two closest SF discs in TNG100, $k_1{=}1.4\rm\,Mpc$, is ${\sim}40\%$ higher than the characteristic radius (1~Mpc) considered in evaluating the isolation parameters in AMIGA \citep[][although a radius of 0.5~Mpc is often considered]{Verley2007a,Argudo-Fernandez2013}. To investigate whether a more conservative or `generous' definition of isolation influences the trend observed in \Cref{fig:jvsm-tng-obs}, we set the search radius $r_v$ to two additional values: $k_1$ and $3k_1$. As shown in the right panel of \Cref{fig:jvsm-tng-fatm}, the change in $r_v$, which greatly affects the sample size of isolated galaxies, does not result in notable differences in the trends of \js.

%----------------------------

\subsection{Gas fraction as a driver of angular momentum content}
With isolation discarded as the sole factor for the high angular momentum of TNG galaxies, an important parameter that could contribute to the increasing \js\ in AMIGA galaxies is the atomic gas fraction. This is evidenced by the trends of \Cref{fig:jvsm-tng-obs} (left panel) for TNG100 subhaloes. In fact, observational studies have found that gas-rich galaxies possess higher specific baryonic angular momentum than gas-normal galaxies \citep[e.g.][]{Lutz2018}. In this context, isolation acts as a proxy parameter that indirectly influences angular momentum through atomic gas content. As established in \citet{Jones2018}, the AMIGA sample has a higher atomic gas content than control samples that were not intentionally selected in low-density environments. This could explain their higher baryonic angular momentum, as measured in \citetalias{Sorgho2024}. A notable point from \Cref{fig:jvsm-tng-fatm} is that the gas content has a subtle impact on \js\ of TNG100 galaxies and mostly affects galaxies with $\Ms{\gtrsim}\,10^{9.5}\Mo$. At $f_{\rm atm}{>}0.2$, they present a trend comparable to that of AMIGA galaxies. This is because, as seen in \Cref{fig:jvsm-amiga}, most low-$f_{\rm atm}$ galaxies lie in this stellar mass range.

\begin{figure*}
    \centering
    \includegraphics[width=\textwidth]{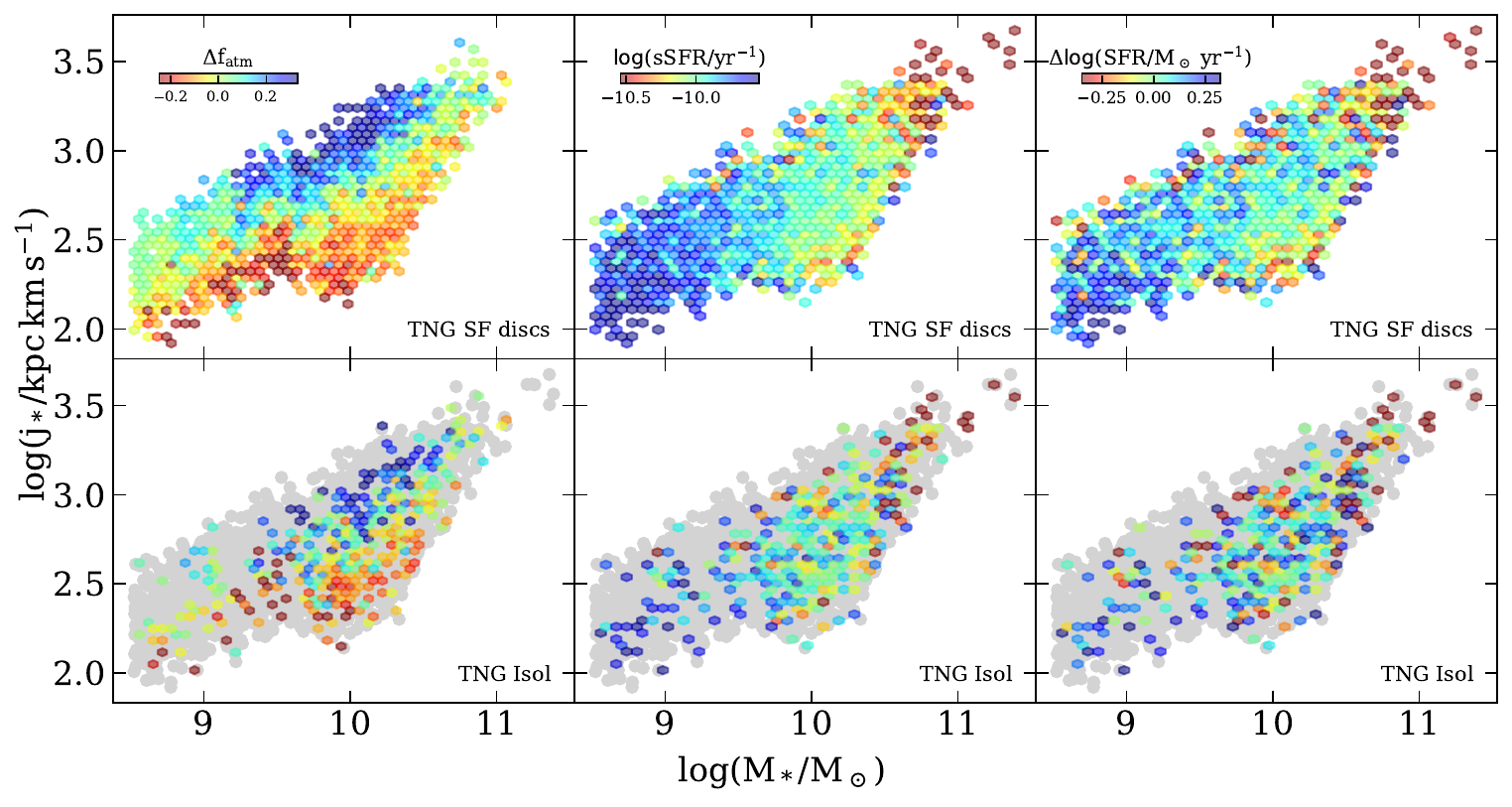}
    \vspace{-20pt}
    \caption{Fall relation of TNG100 subhaloes colour-coded with different parameters: atomic gas excess $\Delta f_{\rm atm}$ (left), sSFR (middle), and deviation from the star formation main sequence (right). Top panels: Star-forming discs. Bottom panels: Most isolated discs.}
    \label{fig:mj-fatm-sfr}
\end{figure*}

Given the galaxy segregation in the TNG100 star-forming sample based on the atomic gas fraction, we employed the gas excess parameter to better evaluate the relationship between \js\ and the gas content. This was determined by subtracting the average value of the gas fraction in a given mass bin from its absolute value. In the left panels of \Cref{fig:mj-fatm-sfr}, we show this excess, $\Delta f_{\rm atm}$, as the third axis of the Fall relation. The dependence of \js\ on the gas content becomes abundantly clear as parallel regions of equal excess gas content are observed from bottom to top in the parameter space of $\Ms{-}$\js. Galaxies on the upper side of \js\ show an excess \hi, while those at the bottom are deficient. We reproduce the Fall relation in the middle panels of the figure, colour-coded by the time-averaged sSFR of the galaxies \citep{Donnari2019a,Pillepich2019}. As detailed in \citet{Donnari2019a}, the SFRs in the TNG model are measured by summing the masses (at birth) of the individual stellar particles formed within a given time frame and averaging them over that time period. The choice of the initial mass over the present-day mass (or the mass at the time of observation) is motivated by the need for consistency with observational indicators, which are generally sensitive to SFRs averaged over timescales. In the present work, we used the SFR averaged over the last 1~Gyr and within a 3D aperture equivalent to twice the stellar half-mass radius. The sSFR is the SFR per unit stellar mass. The trend shows that low-mass galaxies form more stars than higher-mass subhaloes. In fact, the most star-forming galaxies exhibit sSFRs about an order of magnitude higher than those of the least star-forming galaxies. Furthermore, the correlation between \js\ and $\Ms$ implies that galaxies with higher \js\ are the least actively star-forming. Since the galaxies of the sample were selected to be star-forming, a significant fraction is expected to lie on the star-forming main sequence (SFMS). To better investigate any potential correlation of SF activity with \js, we considered the deviation from the SFMS. We employed the definition of the SFMS of \citet{Ma2022} for TNG100:
\begin{equation}
    \log{({\rm SFR_{MS}}/\Mo\,{\rm yr^{-1}})} = 0.83\,\log{(\Ms/\Mo) - 8.32}.
\end{equation}
The deviation from the SFMS is expressed as ${\rm\Delta SFR} = \log{(\rm SFR)} - \log{(\rm SFR_{MS})}$. In the right panels of \Cref{fig:mj-fatm-sfr}, we show the distributions of $\rm\Delta SFR$ in the $\Ms{-}$\js\ parameter space, revealing a similar trend with the sSFR. In particular, we note that galaxies in the lower mass region (and hence lower \js\ region) of the parameter space present a positive deviation, suggesting that they exhibit increased SF activity with respect to their location on the SFMS. This corroborates the findings that galaxies with the highest \js\ values have the most stable discs, therefore exhibiting lower star formation activity \citep[e.g,][]{Swinbank2017}. On the other hand, given the lack of correlation between sSFR and $\Delta f_{\rm atm}$, it contradicts the claim that the anti-correlation between sSFR and \js\ is caused by the fact that galaxies with higher spins (or higher specific angular momentum) experience a less efficient gas infall into their discs \citep[e.g.][]{Lu2022,WangX2022}.

In the lower panels of \Cref{fig:mj-fatm-sfr}, we only include the TNG100 galaxies that satisfy the isolation criteria set in \Cref{sec:data:isol}. In all the probed parameter spaces, they do not seem to prefer any particular region. The isolated subhaloes include both gas-rich and gas-poor galaxies in similar proportions, although the majority lies around $\Delta f_{\rm atm}{\sim}0$. Similarly, the distributions of both sSFR and $\rm\Delta SFR$ of the sample are roughly uniform.

The present analysis does not attempt to isolate variations in the global stellar specific angular momentum that may arise from internal secular processes, such as bar formation and stellar feedback. Rather, we primarily focus on the role of environment, particularly galaxy isolation. A number of studies have suggested a connection between low angular momentum and the presence of stellar bars. This is generally interpreted as evidence that low-angular-momentum discs are more susceptible to the bar instability, rather than that bar formation substantially reduces the global stellar specific angular momentum \citep[e.g.][]{Mo1998,Cervantes-Sodi2013}. Indeed, while stellar bars efficiently redistribute angular momentum through secular evolution, their impact on the galaxy-integrated \js\ is expected to remain modest. Bar-driven torques predominantly redistribute angular momentum from the inner disc to the outer disc and DM halo, profoundly modifying the orbital structure of stars within the bar region \citep{Athanassoula2002,Athanassoula2003,Sellwood2014}. In contrast, the global \js\ is dominated by the extended outer disc, where most of the stellar angular momentum resides owing to its large radial extent. Consequently, even substantial changes to the kinematics of the inner few kiloparsecs are expected to produce only modest variations in the integrated \js. This physical picture is consistent with the weak observational connection between bar presence or strength and global stellar specific angular momentum reported by recent studies \citep[e.g.][]{Romeo2023}.

%--------------------------------------------------------------------

\section{Summary}\label{sec:summary}
In this study, we revisited the relationship between the specific angular momentum and galaxy mass \citep{Fall1983} for isolated galaxies. Our aim was to disentangle possible effects of galaxy number density on the specific stellar angular momentum, \js, in light of the recent study using observed galaxies from the AMIGA sample \citepalias{Sorgho2024}. For this, we used the TNG100 simulation, in which we identified the most isolated subhaloes based on their local density. Conssitent with previous observational and theoretical studies \citep[e.g.][]{Lagos2017,ManceraPina2021a}, we find a dependency of \js\ on the atomic gas fraction, $f_{\rm atm}$. At fixed $\Ms$, galaxies with an excess of atomic gas have higher angular momentum compared with those with lower gas fractions. This is true for both the observed AMIGA galaxies and simulated TNG100 subhaloes. Moreover, we observe a turnover in the median \js\ values of the TNG100 star-forming discs at $\Ms{\sim}10^{10}\,\rm\Mo$, which causes a discrepancy with the observed \js\ values. This turnover region, likely caused by gas-poor subhaloes, is inexistent for galaxies with $f_{\rm atm}{\gtrsim}0.2$. 

Furthermore, a study of the Fall relation for the star-forming discs in TNG100 against their local number density, $\eta_3$, shows no strong correlations with the parameter. This suggests that \js\ does not directly correlate with galaxy isolation. Rather, environment affects angular momentum through the atomic gas content.

\section*{Data availability}
The simulation data used in this work are publicly available through the IllustrisTNG data portal\footnote{\url{https://www.tng-project.org/data}}. All data products derived as part of this study are publicly available via a Zenodo repository\footnote{\url{https://doi.org/10.5281/zenodo.21992329}}. This repository includes all derived data products, as well as the tables and figures presented in this paper. To facilitate the reproducibility of our analysis, the repository also contains the Python scripts and Jupyter notebooks developed for this work. Together, these resources provide access to the data products and analysis pipeline required to reproduce the results presented in this paper.

%--------------------------------------------------------------------

\begin{acknowledgements}
This work used the Spanish Prototype of an SRC \citep[espSRC,][]{Garrido2021} service and support funded by the Ministerio de Ciencia, Innovaci\'on y Universidades (MICIU), by the Junta de Andaluc\'ia, by the European Regional Development Funds (ERDF) and by the European Union NextGenerationEU/PRTR. The espSRC acknowledges financial support from the Agencia Estatal de Investigaci\'on (AEI) through the ``Centre of Excellence Severo Ochoa" award to the Instituto de Astrof\'isica de Andaluc\'ia (IAA-CSIC) (SEV-2017-0709) and together with the authors AS, LVM, RI, MK, BN, SSE and JG from the grant CEX2021-001131-S funded by MICIU/AEI/10.13039/501100011033. AS, LVM, RI, MK and BN acknowledge financial support from the grant PID2021-123930OB-C21 and PID2024-155817OB-I00 funded by MICIU/AEI and by ERDF/EU. MK acknowledges funding through the SAFE -- ``Supporting At-Risk Researchers with Fellowships in Europe'' project, funded by the European Union under Grant Agreement No. 101148426. BN acknowledges financial support from the grant DGP\_POST\_2024\_01021 funded by la Junta de Andaluc\'ia/CUII and by the ESF+.
\end{acknowledgements}

\bibliographystyle{aa}
\bibliography{references}

\begin{appendix}

\section{Impact of orientation on the baryonic Tully-Fisher relation}\label{sec:app:or}
\begin{figure}
    \centering
    \includegraphics[width=\columnwidth]{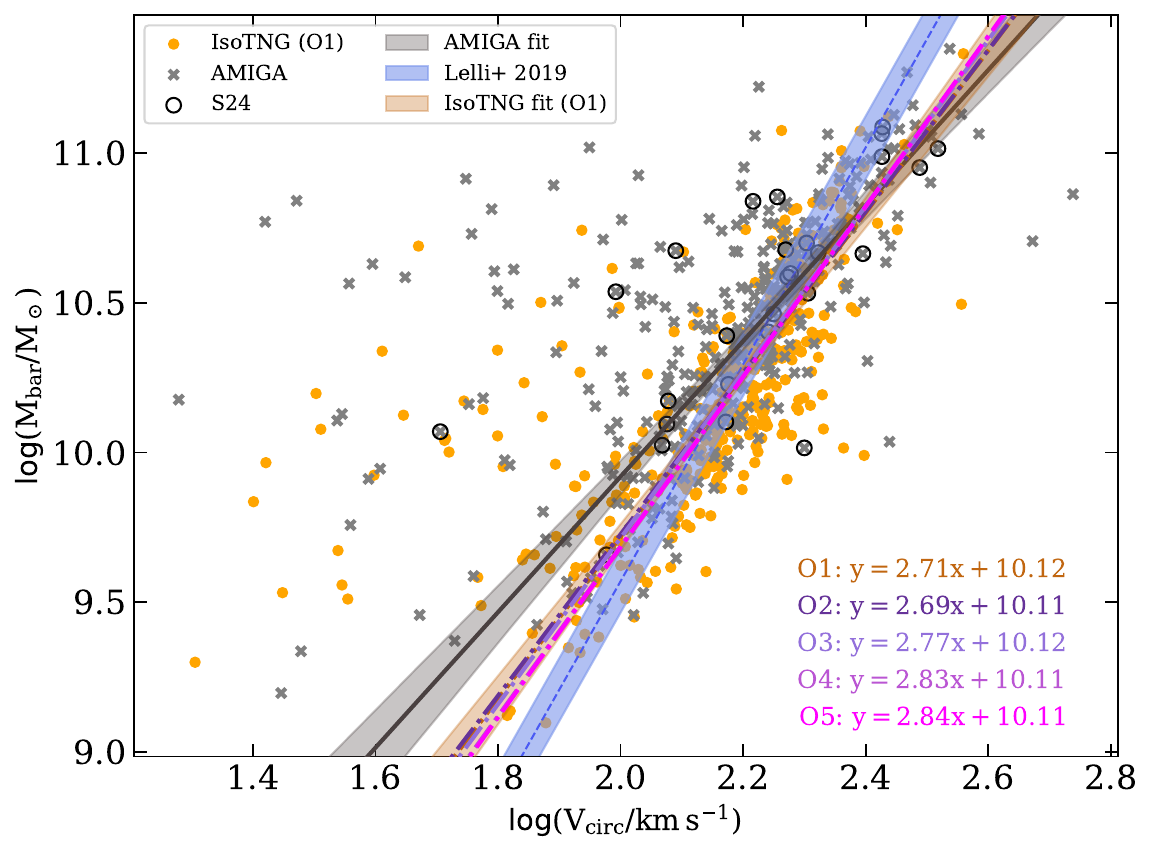}
    \vspace{-20pt}
    \caption{bTFr of isolated TNG100 galaxies for different SKIRT orientations.}
    \label{fig:btfr-or}
\end{figure}

In \Cref{sec:res:btfr} we arbitrarily selected the SKIRT orientation O1 of the generated galaxy profiles. As mentioned therein, the choice of the orientation is arbitrary and should not affect the shape of the bTFr. In \Cref{fig:btfr-or}, we reproduce the bTFr at different orientations. We note that the relations for all orientations O2 to O5 lie within the standard deviation of that of O1, suggesting that no significant deviation occurs when one switches from one orientation to another. This is expected as, in theory, the distribution of the galaxies in the simulation box does not favour any particular direction.

\end{appendix}

\end{document}